\documentclass[sigconf]{acmart}
\theoremstyle{definition}

\usepackage{subfig}
\usepackage{graphics}
\usepackage{float}
\usepackage{natbib}
\usepackage{graphicx}
\usepackage{caption}
\usepackage{color}
\usepackage{changepage}
\usepackage{graphicx}
\usepackage{amsmath}
\usepackage[linesnumbered,ruled]{algorithm2e}
\pdfpageattr {/Group << /S /Transparency /I true /CS /DeviceRGB>>}

\AtBeginDocument{%
  \providecommand\BibTeX{{%
    \normalfont B\kern-0.5em{\scshape i\kern-0.25em b}\kern-0.8em\TeX}}}

\setcopyright{none}
\renewcommand\footnotetextcopyrightpermission[1]{} 

\setcopyright{none}
\renewcommand\footnotetextcopyrightpermission[1]{} 

\begin{document}
\title{Quantitative Stock Investment by Routing Uncertainty-Aware Trading Experts: A Multi-Task Learning Approach}

\author{Shuo Sun\qquad Rundong Wang \qquad Bo An\\ Nanyang Technological University}

\renewcommand{\shortauthors}{}

\begin{abstract}
Quantitative investment is a fundamental financial task that highly relies on accurate stock prediction and profitable investment decision making. Despite recent advances in deep learning (DL) have shown stellar performance on capturing trading opportunities in the stochastic stock market, we observe that the performance of existing DL methods is sensitive to random seeds and network initialization. To design more profitable DL methods, we analyze this phenomenon and find two major limitations of existing works.  
First, there is a noticeable gap between accurate financial predictions and profitable investment strategies. Second, investment decisions are made based on only one individual predictor without consideration of model uncertainty, which is inconsistent with the workflow in real-world trading firms. To tackle these two limitations, we first reformulate quantitative investment as a multi-task learning problem. Later on, we propose AlphaMix, a novel two-stage mixture-of-experts (MoE) framework for quantitative investment to mimic the efficient bottom-up trading strategy design workflow of successful trading firms. In Stage one, multiple independent trading experts are jointly optimized with an individual uncertainty-aware loss function. In Stage two, we train neural routers (corresponding to the role of a portfolio manager) to dynamically deploy these experts on an as-needed basis. AlphaMix is also a universal framework that is applicable to various backbone network architectures with consistent performance gains. Through extensive experiments on long-term real-world data spanning over five years on two of the most influential financial markets (US and China), we demonstrate that AlphaMix significantly outperforms many state-of-the-art baselines in terms of four financial criteria. Furthermore, we analyze the contributions of each component in AlphaMix with a series of exploratory and ablative studies. Our data and source code are publicly available \footnote{https://anonymous.4open.science/r/AlphaMix-C767/}.
\end{abstract}

\maketitle
\pagestyle{plain}
\section{Introduction}
The stock market, a financial ecosystem involving over \$90 trillion\footnote{https://data.worldbank.org/indicator/CM.MKT.LCAP.CD/} market capitalization globally in 2020, is one of the most popular channels for investors to pursue desirable financial assets and achieve investment goals. As the efficient market hypothesis (EMH) \cite{fama1970efficient} claims, forecasting stock prices accurately and making profitable investment decisions are extremely challenging tasks due to the high volatility and noisy nature of the financial markets \cite{adam2016stock}. In the last decade, deep learning (DL) methods have become an appealing direction for stock prediction and quantitative investment \cite{cavalcante2016computational} owing to its ability to learn insightful market representations in an end-to-end manner. Most existing DL methods adopt various advanced network architectures (e.g., LSTM \cite{nelson2017stock}, Attention \cite{qin2017dual} and Transformer \cite{ding2020hierarchical}) to extract meaningful features from heterogeneous financial data such as 
fundamental factors \cite{chauhan2020uncertainty}, 
economics news \cite{hu2018listening}, social media \cite{xu2018stock} and investment behaviours \cite{chen2019investment}.

\begin{figure}[t]
  \centering
  \subfloat[{}\label{fig:confidence}]{%
      \includegraphics[width=0.23\textwidth]{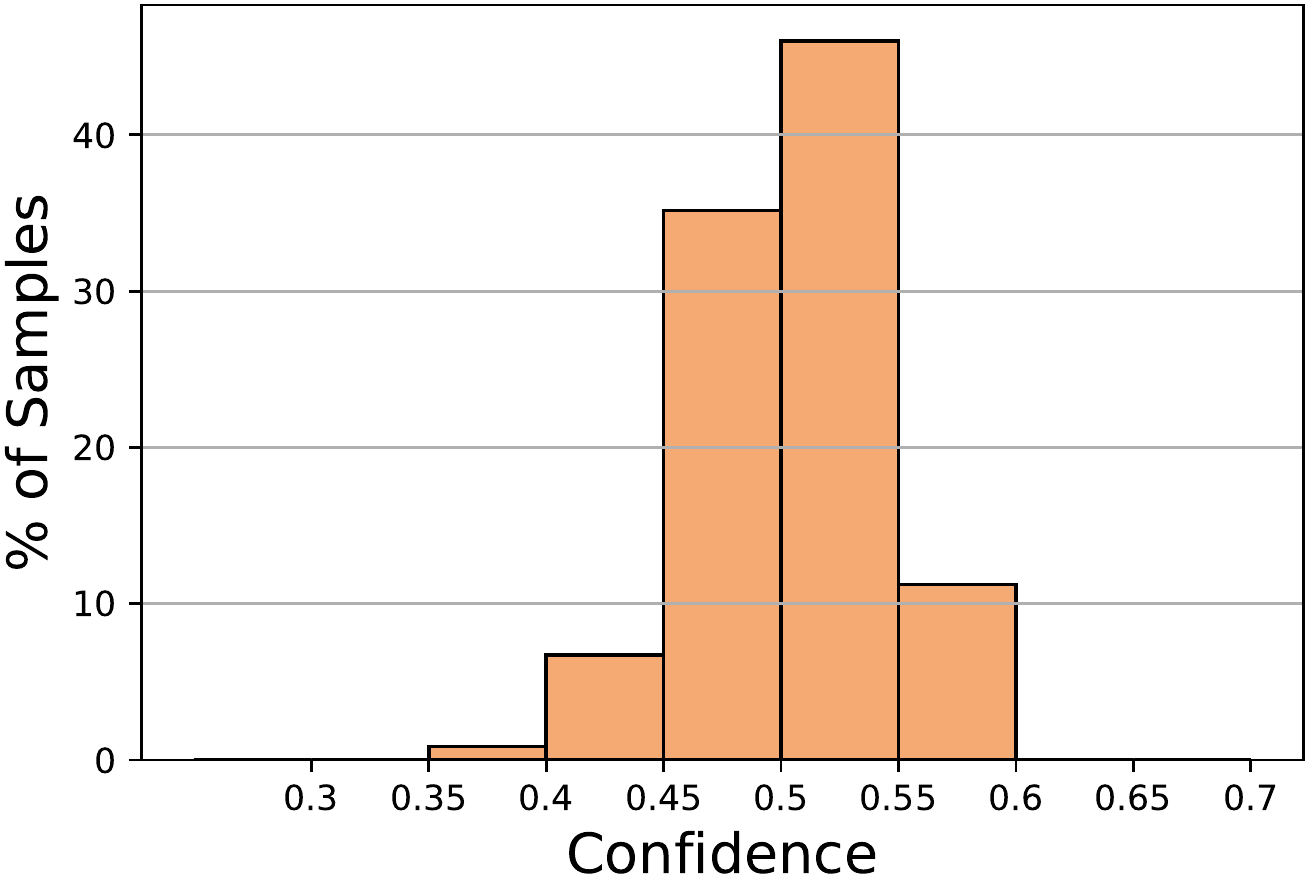}
     }
     \hfill
 \subfloat[{}\label{fig:variance}]{%
      \includegraphics[width=0.23\textwidth]{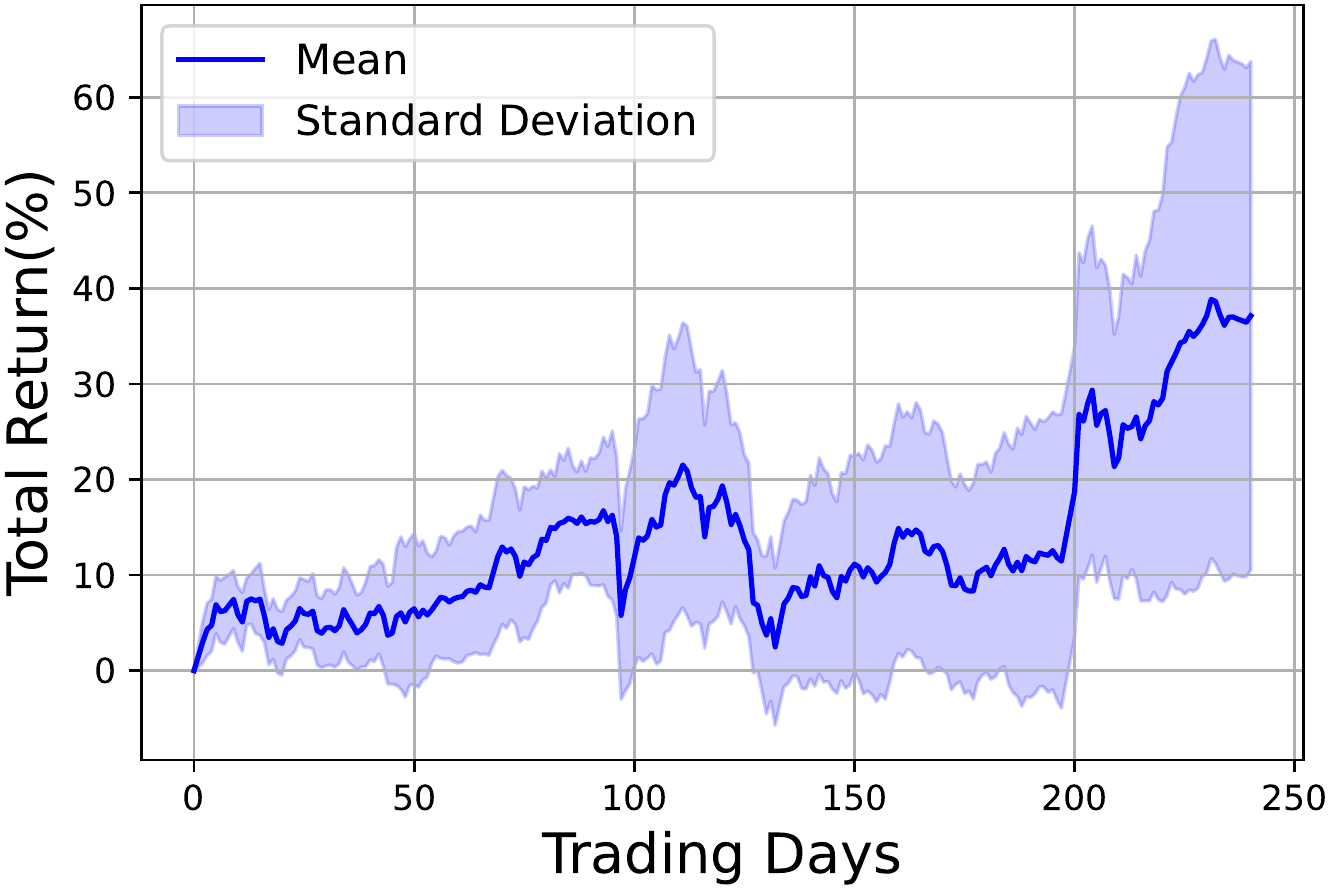}
     }

     \caption{(a) Limitation 2: Confidence histogram of GRU to show single deep learning predictor's high uncertainty on stock movement prediction. (b) Sensitivity: Trading days vs. total return curve of GRU to show its performance is sensitive to random seeds and network initialization. }
     \label{fig:trading_behavior}
\end{figure}

To apply deep learning models for quantitative investment, the vast majority of existing methods firstly adopt stock prediction as a supervised learning task. Observable market information (e.g., historical price) is used as the feature vector and the future price movement direction or return rate is applied as the target for classification or regression task, respectively. A predictor is optimized to capture the underlying pattern of financial markets by minimizing the corresponding classification or regression loss function. Later on, the trained predictor is applied to new unseen test sets for prediction. Finally, investment strategies are constructed by picking stocks with top predicted price rising probability or return rate. However, the performance of previous methods is sensitive to random seeds and network initialization \cite{ballings2015evaluating}. We conduct experiments to further demonstrate the sensitivity. We train a popular two-layer GRU \cite{shen2018deep} model for stock movement prediction and generate investment decisions with the classic top-4 buy \& hold trading strategy following \cite{yoo2021accurate,lin2021learning} with 10 independent runs. The trading days vs. total return curve (mean and standard deviation) of the GRU model is reported in Figure \ref{fig:variance}. Even though it achieves a satisfactory total return (almost 40\%), the large performance standard deviation (shaded area) is still a major concern for high-stake decision making problems such as quantitative investment. We analyze the potential reasons of this phenomenon as follows:

\begin{figure}[t]
\begin{center}
\includegraphics[width=0.48\textwidth]{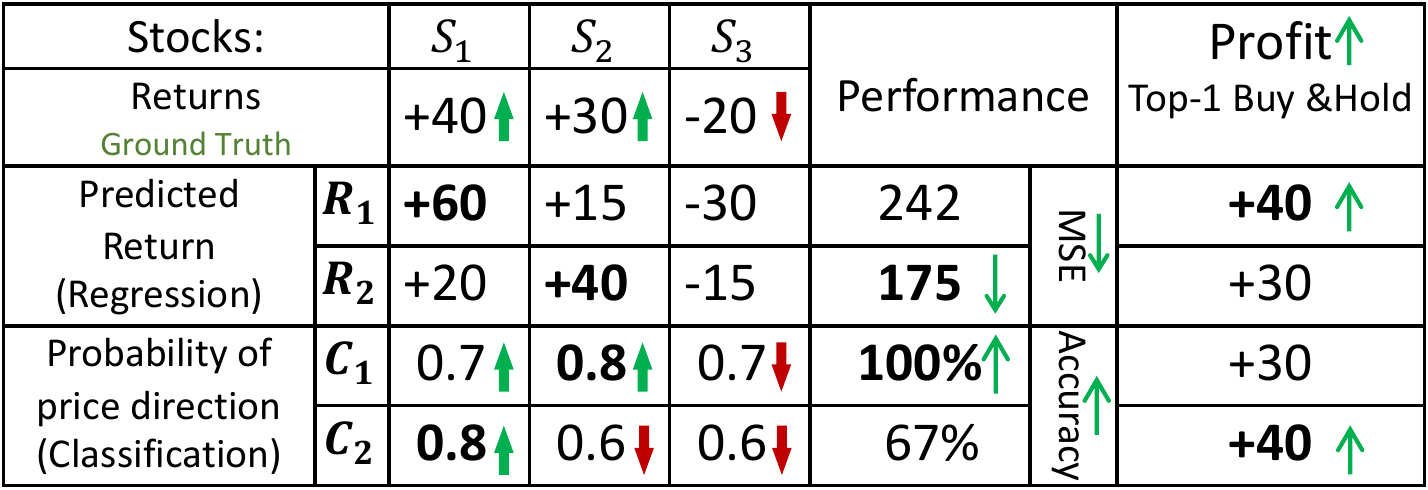}
\end{center}
\caption{Limitation 1: There is a noticeable gap between accurate financial prediction and profitable investment decisions. More accurate prediction $\mathbf{R_2( MSE\downarrow),C_1(Acc.\uparrow)}$ can sometimes lead to lower profit (+30 vs. +40) than less accurate prediction $\mathbf{R_1(MSE\uparrow), C_2(Acc.\downarrow)}$. }
\label{fig:gap}
\end{figure}

First, the widely used supervised learning pipeline is not directly optimizing profits (the key objective for quantitative investment), which leads a noticeable gap between accurate financial predictions and profitable trading decisions \cite{sawhney2021stock}. As illustrated in Figure \ref{fig:gap}, $\mathbf{R_2}$ and \(\mathbf{C_1}\) achieve better prediction performance (lower loss or higher accuracy) than \(\mathbf{R_1}\) and \(\mathbf{C_2}\). However, while making investment decisions following the top-1 buy \& hold strategy, \(\mathbf{R_2}\) and \(\mathbf{C_1}\) pick stock \(S_2\) and get lower profit than \(\mathbf{R_1}\) and \(\mathbf{C_2}\) that pick stock \(S_1\) (+30 vs. +40).     
Considering the real-world trading scenario, experienced human experts evaluate stocks by judging the probability of price rising (classification) and the scale of price value change (regression) as two complementary tasks to support investment decision making. This hints us to reformulate quantitative investment as a multi-task learning task to narrow down the gap by learning price movement and return rates simultaneously.

\begin{figure}[t]
\begin{center}
\includegraphics[width=0.48\textwidth]{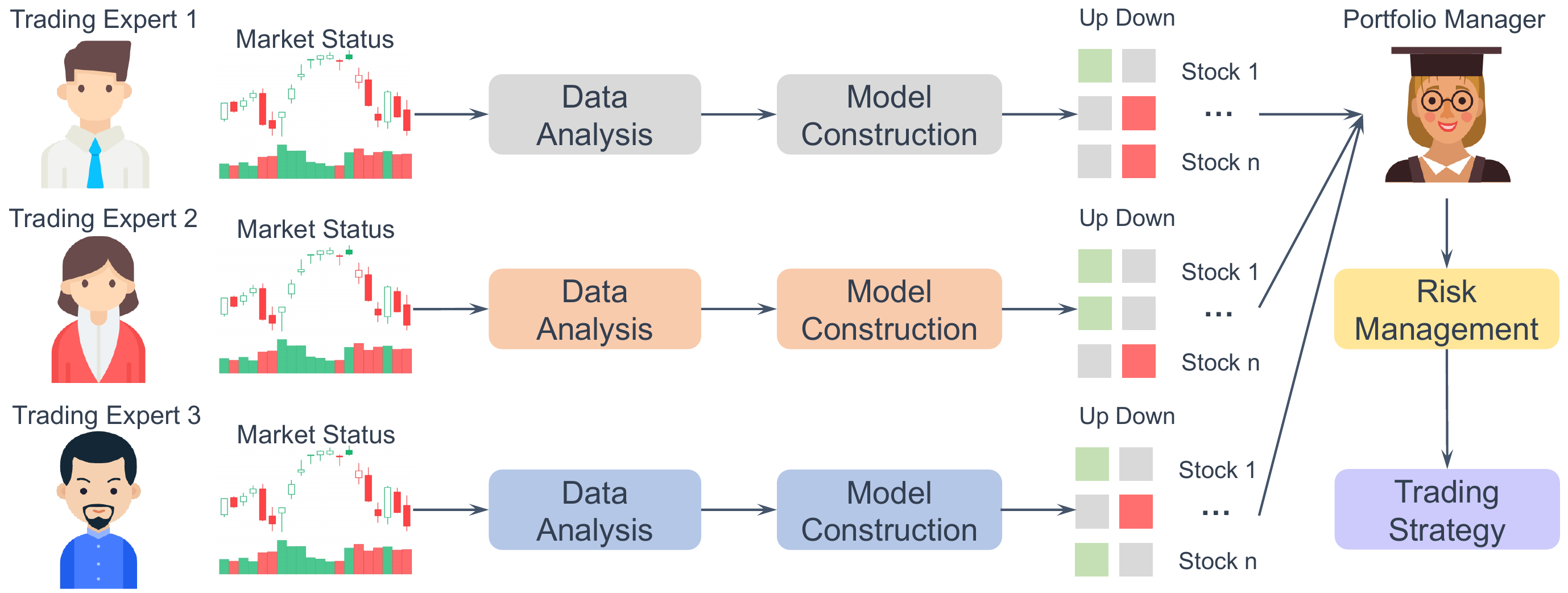}
\end{center}
\caption{Overview of the bottom-up hierarchical trading strategy design workflow in real-world trading companies.}
\label{fig:pipeline}
\end{figure}

Second, existing DL methods make investment decisions based on only one individual predictor without consideration of model uncertainty. In Figure \ref{fig:confidence}, we observe that the confidence score (largest logits after softmax) for stock movement prediction (binary classification) is quite low (around 0.5 for most samples) that indicates one single DL predictor is not enough due to the high uncertainty. In most real-world trading firms (Figure \ref{fig:pipeline}), multiple trading experts firstly share all information and have a discussion to achieve an agreement of the current market status. Then, they conduct data analysis and build their own models independently based on personal trading style and risk tolerance. Finally, a portfolio manager summarizes their results, conducts risk management (uncertainty estimation in finance), and makes the final investment decisions. This bottom-up hierarchical workflow with uncertainty estimation plays a key role for designing robust and profitable trading strategies. Inspired by it, we propose AlphaMix, a novel two-stage mixture-of-experts (MoE) framework for quantitative investment. In Stage one, AlphaMix trains multiple independent trading experts with an individual uncertainty-aware loss function. In Stage two, we achieve better performance and computational efficiency by training neural routers to dynamically pick experts on an as-needed basis. The main contributions of this work are three-fold:
\begin{itemize}
    \item We conduct experiments to analyze why the performance of existing DL methods is sensitive and reformulate quantitative investment as a multi-task learning problem following the analysis procedure of human experts.
    \item To the best of our knowledge, AlphaMix is the first universal mixture-of-experts framework for quantitative investment. In AlphaMix, we firstly optimize multiple independent trading experts with an individual uncertainty-aware loss function and then train neural routers to dynamically pick these experts on an as-needed basis.
    \item Through experiments on real-world data of two influential stock markets spanning over 5 years, we show that AlphaMix significantly outperforms many state-of-the-art baseline methods in terms of four financial criteria and demonstrate AlphaMix's practical applicability to quantitative stock investment with a series of exploratory and ablative studies.
\end{itemize}

\section{Related Work}
\subsection{Stock Prediction with Deep Learning}
Extensive deep learning methods have been proposed for stock prediction, which can be generally categorised into: 1) recurrent neural networks (RNN), 2) non-recurrent neural networks (NRNN), and 3)models with alternative data sources instead of prices.

\noindent
\textbf{RNN models.}
RNN-based models are popular for stock prediction since they are specifically designed to capture temporal patterns in sequential data. \citet{nelson2017stock} show that LSTM and GRU outperform many baseline methods. \citet{wang19} combine attention mechanism with LSTM to model correlated time steps. To improve the robustness of LSTM, \citet{feng2018enhancing} apply adversarial training techniques for stock prediction. \citet{zhang2017stock} propose a novel State Frequency Memory (SFM) recurrent network with Discrete Fourier Transform (DFT) to discover multi-frequency patterns in stock markets.

\noindent
\textbf{NRNN models.}
There are many DL approaches that make stock prediction without using RNN-based models. \citet{liu2020multi} introduce a multi-scale two-way neural network (NN) to predict the stock trend.
\citet{ding2020hierarchical} propose a hierarchical Gaussian transformer-based model for stock movement prediction. Another direction of work employs graph-based DL models to model pair-wise relations between stocks. \citet{feng2019temporal} enhance graph convolutional networks (GCNs) with temporal convolutions for mining inter-stock relations. \citet{sawhneydeep} focus on stock industry data and links between company CEOs. 


\noindent
\textbf{Models with alternative data.}
To further show the potential of DL methods, the usage of additional data sources is another direction for trading signal discovery. \citet{xu2018stock} propose a variational autoencoder architecture to extract latent information from tweets. \citet{chen2019investment} enhance trading strategy design with the investment behaviors of professional fund managers. Other data sources such as economics news \cite{hu2018listening} and earning calls \cite{sawhney2020voltage} are also used to improve the prediction performance. 

However, this line of work focuses on designing one powerful DL model to extract meaningful market features. These models usually involve millions of parameters and are hard to train in practice. Our AlphaMix proposes a new direction to make investment decisions with multiple models and significantly outperforms existing SOTA methods with simple NN backbones (e.g., MLP).

\subsection{Ensemble Learning in Stock Market }
To better understand the stochastic nature of the stock market, ensemble learning is an effective way to enhance model robustness on financial data \cite{nti2020comprehensive}. Bagging methods \cite{breiman1996bagging}, which build diversified individual sub-models separately, are the mainstream for financial prediction. For example, \citet{xiang2006predicting} explore the performance of different base models to construct sub-models for bagging. \citet{liang2012stock} and \citet{zhai2010hybrid} build sub-models by dynamically selecting financial data based on time periods and market environments, respectively. \citet{zhang2020doubleensemble} propose DoubleEnsemble, a unified framework that ensembles sub-models with sample reweighting and feature selection. Other ensemble methods such as boosting \cite{long2010random} is currently not widely applied for financial prediction due to their tendency to overfit to the noise in the training data \cite{saxena2019data}.

However, existing works simply apply ensemble approaches by training one model each time with different initialization and combining the prediction of all models in the inference period. Our AlphaMix is a non-traditional ensemble method. It enables to train multiple trading experts simultaneously with uncertainty estimation and dynamically deploys experts on an as-needed basis.

\subsection{Modeling Financial Uncertainty}
Modeling uncertainty (aka. risk) in the financial markets has been a popular topic in the finance community for decades. \citet{artzner1999coherent} present and justify a set of four desirable properties for measures of uncertainty, and call the measures satisfying these properties "coherent". \citet{rockafellar2002conditional} propose conditional value-at-risk (CVaR) as a coherent measure of risk to provide short-cuts for many financial optimization problem. \citet{acharya2017measuring} present an economic model to measure each financial institution's contribution to systemic risk as its systemic expected shortfall (SES). \citet{jurado2015measuring} exploit a data rich environment and define common volatility as a direct econometric estimate of time-varying macroeconomic uncertainty. 

However, there is very limited work focusing on modeling financial uncertainty in machine learning models. \citet{sawhney2021modeling} propose a multi-modal time-aware hierarchical entropy-based curriculum learning framework to model financial uncertainty. In our AlphaMix, we design a novel uncertainty-aware loss function to model financial uncertainty in another way.

\subsection{Deep Learning with Mixture of Experts}
Ever since its introduction more than two decades ago \cite{jacobs1991adaptive}, the mixture-of-experts (MoE) approach has been the subject of much research. \citet{eigen2013learning} extend MoE into deep learning by stacking two layers of mixture of experts with a trainable weighted gating network. \citet{shazeer2017outrageously} present a sparsely-gated MoE layer to enable training extremely large number of experts. \citet{ma2018modeling} propose a multi-gate MoE framework to model task relationships in multi-task learning. MoE models have achieved stellar performance in many challenging fields such as computer vision \cite{ruiz2021scaling,wang2020long,wang2020deep} and natural language processing \cite{shazeer2017outrageously,du2021glam}.

However, even though MoE methods have been successful in many fields, there is a lack of MoE frameworks for the stochastic financial market. AlphaMix is the first MoE framework to fill this gap with a novel two-stage training mechanism to mimic the efficient hierarchical bottom-up trading strategy design workflow in real-world trading firms.

\section{Problem Formulation}
We formulate quantitative investment as a multi-task learning problem by jointly learning to predict stock movement direction and return rate. Let \(\mathcal{S} = \{s_1,s_2,...,s_N\}\) denote the set of N stocks, where for each stock \(s_i \in \mathcal{S}\) on trading day \(t\), there is a corresponding closing price \(p_t^i\) and a feature vector \({x}_t^i\).

\noindent
\textbf{Formalizing stock movement prediction:} Following \cite{xu2018stock}, we define stock movement prediction with label \(y^i_{[t,t+\tau]}\) over a period of \(\tau\) days as a binary classification task. Here, we define the label using the closing price of a given stock  \(p_i^t\) that can either rise or fall on day \(t+\tau\) compared to day \(t\) as:
\begin{equation}
y^i_{[t, t+\tau]}=\begin{cases}
 1,\quad p^i_{t+\tau}>p_t^{i}\\0,\quad p^i_{t+\tau}\le p_t^{i}
\end{cases}
\end{equation}

\noindent
\textbf{Formalizing stock return prediction:} Following \cite{zhang2017stock}, we define stock return prediction over a period of \(\tau\) days as a regression task with target return as:
\begin{equation}
    r_{[t, t+\tau]}^i = (p^i_{t+\tau}-p^i_{t})/{p_t^{i}} 
\end{equation}

\noindent
\textbf{Multi-task learning objective:} In the prediction phase, our main objective is to simultaneously predict stock movement \( y^i_{[t, t+\tau]}\) and stock return \(r^i_{[t, t+\tau]}\) using a sequence of feature vectors \(({x}_{t-k}^i,...,{x}_t^i)\) with length \(k+1\). In the following of this work, we ignore the stock index and time step, and use \(y\), \(r\) and \({x}\) to denote stock movement, stock returns, and feature vectors, respectively, for simplicity. 


\noindent
\textbf{Trading strategy for profit generation:} To evaluate the performance of our approach in a real-world scenario, we extend the widely used daily top \(k\) buy \& hold trading strategy \cite{li2016quantitative,yoo2021accurate,lin2021learning} with a uncertainty estimation heuristic (corresponding to the risk management procedure in finance). More details are available in Section 4.4. 

The advantages of this multi-task learning formulation are two-fold: 1) Deep learning models can learn better market embeddings by jointly learning these two complementary tasks. 2) Predictions of stock price from multiple perspectives (movement direction and return rate) can shed light on more profitable investment decisions. 



\section{Methodology}

\begin{figure*}[t]
\begin{center}
\includegraphics[width=0.98\textwidth]{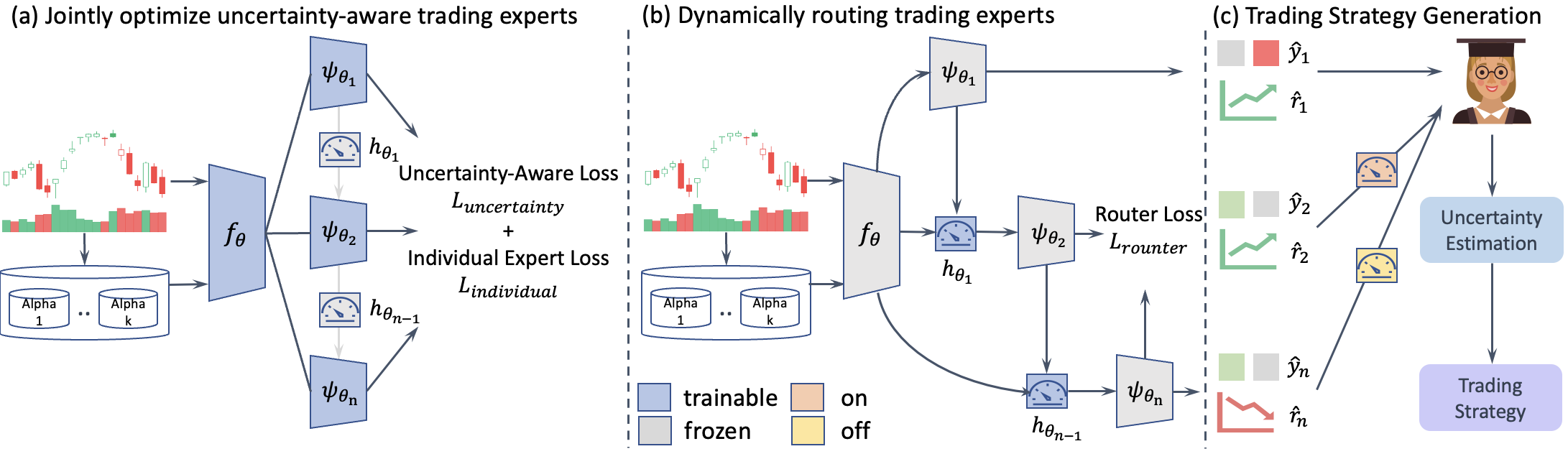}
\caption{A high-level overview of the AlphaMix framwork. a) We first jointly optimize multiple trading experts with an individual uncertainty-aware loss function. b) We then train neural routers to dynamically allocate trading experts on an as-need basis. c) We generate profitable trading strategies with a heuristic estimation of uncertainty.}
\label{fig:alphamix}
\end{center}
\end{figure*}

In this section, we propose AlphaMix (Figure. \ref{fig:alphamix}), a novel two-stage mixture-of-expert framework, to generate highly profitable quantitative investment strategies. Specifically, AlphaMix trains multiple individual uncertainty-aware experts in Stage one and then dynamically routes these experts to further improve the performance and cut down the computational cost in Stage two. In the inference period, AlphaMix generates profitable trading strategies with uncertainty estimation. 

\subsection{Training Multiple Individual Experts}
Considering a real-world trading company with \(n\) independent trading experts trying to extract insightful features based on financial market observations, they firstly share all information together and have a discussion to achieve an agreement of the current market status. Later on, they make independent investment decisions based on personal trading style and risk tolerance. In AlphaMix, we mimic this procedure with the same hard parameter sharing backbone \(f_{\theta}\) following the common practice in MoE systems \cite{wang2020long}. The backbone that shares early layers of the network tends to learn generic low-level market features. Each trading expert \(i\) retains independent later NN layers with parameter \(\psi_{\theta_i}\) (for   \(i=1,...,n)\) to generate personal predictions of the market. AlphaMix is a general framework that fits for any NN architecture (e.g., MLP, LSTM and GRU). The motivation here is to learn a shared embedding of the stock market and make individual trading decisions based on it.

\noindent
\textbf{Individual Expert Loss.}
We firstly feed the feature vector\footnote{\label{note1}Unless otherwise stated, we use \(x\), \(y\), \(r\) to denote the sequential feature vectors \(({x}_{t-k}^i,...,{x}_t^i)\), next day stock movement $y^i_{[t, t+1]}$ and return rate $r^i_{[t, t+1]}$  of stock \(i\) on time \(t\) in the following of this work for simplicity.} \({x}\) into AlphaMix and get predicted stock movement \(\hat{y_i}\)  and return \(\hat{r_i}\) of expert \(i\)  as:
\begin{equation}
    \psi_{\theta_i}(f_\theta(x)) = (\hat{y_i} ,\hat{r_i} )
\end{equation}
To combine the loss of multiple experts, one way is to use the aggregated classification logits and the average prediction value for regression. We define the aggregated loss as collaborated: 
\begin{gather}
   L^{classify}_{collaborative}(x,y)=L_{classify}\left( \frac{1}{n}\sum\nolimits_{i=1}^{n}\hat{y_i},y \right)  \\
   L^{regress}_{collaborative}(x,r)=L_{regress}\left( \frac{1}{n}\sum\nolimits_{i=1}^{n} \hat{r_i}, r \right)
\end{gather}
where $L_{classify}$ denotes any classification loss (e.g., cross-entropy) and $L_{regress}$ denotes any regression loss (e.g., mean square error).




However, collaborative loss leads to correlated decisions instead of complementary experts in our preliminary experiments. To discourage correlations, we require each expert to do the job well by itself, where we apply an individual loss that significantly outperforms collaborative loss in practice with a hyperparameter $\lambda$ as a relative importance weight:
\begin{gather}
    L^{classify}_{individual}(x,y)= \frac{1}{n}\sum\nolimits_{i=1}^{n} L_{classify}(\hat{y_i}, y) \\
    L^{regress}_{individual}(x,r)= \frac{1}{n}\sum\nolimits_{i=1}^{n}L_{regress}(\hat{r_i},r ) \\
    L_{individual} = L^{classify}_{individual}(x,y) + \lambda \cdot L^{regress}_{individual}(x,r) 
\end{gather}

\subsection{Joint Uncertainty-Aware Optimization}
\noindent
\textbf{Uncertainty-Aware Loss.}
The individual loss function and random initialization lead to multiple diversified trading experts. For low signal-to-noise financial data, we add two regularization terms for stock movement classification and return regression to encourage trading decisions with aware of uncertainty from multiple experts. First, we define the variation ratio \cite{gal2016uncertainty}, which is a measure of statistical dispersion widely used to quantify the uncertainty of classification models as:
\begin{equation}
    L_{vr} = min(1-\frac{n_{up}}{n}, \frac{n_{up}}{n})
\end{equation}
where \(n\) is the total number of experts and \(n_{up}\) is the number of experts \(i\) that predict the stock price will go up in the next day (equal to $\hat{y_i}=1$). \(L_{vr}\) gets the lowest uncertainty 0 when all experts have the same opinion on the stock movement direction. 

Second, we apply the volatility \cite{jurado2015measuring}, a widely used coherent measure of risk, to quantify the uncertainty of the stock return regression task. It is defined as the variance of the predicted return rate from \(n\) experts: 
\begin{equation}
    L_{vol} = \sigma([\hat{r_1},\hat{r_2},...,\hat{r_n}])  
\end{equation}
Finally, we define the uncertainty loss function as:
\begin{equation}
    L_{uncertainty} = w_1 \cdot L_{vr} + w_2 \cdot L_{vol}
\end{equation}
It is worth noting that there is no need to tune \(w_1\) and \(w_2\). AlphaMix achieves good performance when \(w_1=w_2=1\) in practice.

For our \(n\) experts \((\theta_1,...,\theta_n)\) with a shared backbone \(\theta\), we jointly optimize the sum of the individual loss and uncertainty loss as the total loss function:
\begin{equation}
    L_{}(x,y) =\sum\nolimits_{i=1}^{n}(L_{individual}(x,y;\theta_i)+L_{uncertainty}(x,y;\theta_i))
\end{equation}
Since these loss terms are completely symmetrical with respect to each other, the \(n\) experts learned at Stage 1 are equally important and aware of the overall uncertainty from each other.

\subsection{Dynamically Routing Multiple Experts}
To further improve the performance and computation efficiency, we train \(n-1\) routers with parameters \(h_{\theta_1},...,h_{\theta_{n-1}}\) at Stage 2 to dynamically deploy these (arbitrarily ordered) experts sequentially on an as-needed basis. When the mean logits and returns of the first \(k\)-th expert provide the correct prediction (\(\hat{y}=y\) and \(sign(\hat{r})=sign(r)\)), which indicates the first \(k\) experts are enough, the router should ideally switch off the \(k+1\) expert with \(y_{on} =0\). Otherwise, the router should turn on the \(k+1\) expert to explore more possibility with \(y_{on}=1\). In practice, we build a binary classifier with two fully connected layers to learn each router. Each of the \(n-1\) routers for \(n\) experts has a shared component to reduce feature dimensions and an individual component to make decisions. Specifically, we normalize the market feature $f_{\theta}(x)$ and reduce the embedding dimension (8 in our experiments) by a fully connected layer \(W_1\), which is shared with all routers, followed by LeakyReLU, and then concatenate it with the mean logits  \(\hat{y}^k_{mean} = \frac{1}{k}{\sum_{i=1}^{k}\hat{y_i}} \) and mean return \(\hat{r}^k_{mean} = \frac{1}{k}{\sum_{i=1}^{k}\hat{r_i}} \) of the first \(k\) experts. Furthermore, we project it to a scalar by \(W_2^{(k)}\) which is independent between routers, and finally apply Sigmoid function $S(x) = \frac{1}{1+e^{-x} } $ to get a continuous activation value \(v(x) \in [0,1]\):

\begin{equation}
   v(x) = S\left (W_2^{(k)}\begin{bmatrix}
 LeakyRelu(W_1 \frac{f_\theta(x)}{\left \| f_\theta(x) \right \| } )\\
\\\ Concat[\hat{y}^k_{mean},\hat{r}^k_{mean}]
\end{bmatrix}\right ) 
\end{equation}

The routers are optimized with a weighted variant of binary cross-entropy loss:
\begin{equation}
    L_{router} = -\omega. y_{on}log(v(x))-(1-y_{on})log(1-v(x))
\end{equation}
where \(\omega\) controls the easiness to switch on the routers. We find \(\omega=1.7\) to be a good trade-off between performance and computational cost for all datasets. We present the whole training procedure of AlphaMix in Algorithm \ref{alg:alphamix}. 

\begin{algorithm}[th]
    \SetKwRepeat{REPEAT}{REPEAT}{REPEAT}
    \KwIn{shared backbone $f_{\theta}$, $\psi_{\theta_1},...,\psi_{\theta_n}$ for $n$ experts, $h_{\theta_1},...,h_{\theta_{n-1}}$ for $n-1$ routers. hyperparameter: learning rate \(\alpha\), batch size \(b\), coefficients for different losses: \(\lambda, w_1, w_2\)}
    \KwOut{$f^*_{\theta}$,$\psi^*_{\theta_1},...,\psi^*_{\theta_n}$,$h^*_{\theta_1},...,h^*_{\theta_{n-1}}$}
    \textbf{Stage 1:} Training Multiple Uncertainty-Aware Experts \\
    \Repeat{convergence}{Set a batch of data samples \((x,y,r)\) \\ Compute individual loss $L_{individual}$ \hfill $\rhd$ Eq. (8)\\ Compute variation ratio $L_{vr}$ \hfill $\rhd$ Eq. (9) \\ Compute volatility $L_{vol}$ \hfill $\rhd$ Eq. (10) \\ Compute uncertainty loss $L_{uncertainty}$ \hfill $\rhd$ Eq. (11) \\ Compute the total loss $L$ \hfill $\rhd$ Eq. (12) \\ Update $f_{\theta}$, $\psi_{\theta_1},...,\psi_{\theta_n}$ via gradient descent algorithms }

    \textbf{Stage 2:} Dynamically Routing Experts \\
    \Repeat{convergence}{Set a batch of data samples \((x,y_{on})\) \\ Compute activation \(v(x)\) \hfill $\rhd$ Eq. (13)\\
    Compute router loss $L_{router}$ \hfill $\rhd$ Eq. (14)\\
    Update $h_{\theta_1},...,h_{\theta_{n-1}}$ via gradient descent algorithms}
    
    \caption{Two-Stage Training of AlphaMix}
    \label{alg:alphamix} 
\end{algorithm}

\subsection{Investment with Uncertainty Estimation}
While generating investment decisions, we extend the widely used daily top \(k\) buy \& hold trading strategy \cite{li2016quantitative,yoo2021accurate,lin2021learning} with a uncertainty estimation heuristic (corresponding to the role of risk management in finance). That is, investors compute the mean values of activated experts as the predicted stock movement and the return rates of the next day for each stock when the stock market closes on trading day \(t-1\). We define the predictions with consistent sign on both tasks (e.g., rising  with positive return) as "certain". Then, investors buy the top \(k\) stocks with the highest rising probability from the "certain" stocks at the beginning of the trading day \(t\) and sell the bought stocks at the end of the trading day \(t\). 



\section{Experiment Setup}
\subsection{Datasets and Features}
To conduct a comprehensive evaluation of AlphaMix, we evaluate it on \(\mathit{two}\) real-world datasets from \(\mathit{US}\) and \(\mathit{Chinese}\) stock markets spanning over \(\mathit{five}\) years. We summarize statistics of the two datasets in Table \ref{tab:dataset} and further elaborate on them as follows:

\(\mathit{ACL18}\) \cite{xu2018stock} is a widely used public dataset collected using Google Finance\footnote{Google Finance: https://www.google.com/finance}, which contains historical stock prices of 87 US stocks with high liquidity from Nasdaq and NYSE exchanges. 

\(\mathit{SZ50}\) \cite{yang2020qlib} is a popular dataset collected from Wind\footnote{Wind: https://www.wind.com.cn/}, which contains 4-year historical prices of 47 influential stocks with top capitalization from the Shanghai exchange.

For a fair comparison, we use the same training, validation, and test splits and strictly follow the data preprocessing pipeline as in \cite{xu2018stock}. Furthermore, we generate 11 temporal features as shown in Table \ref{tab:feature} to describe the stock markets following \cite{yoo2021accurate}. $z_{open}$, $z_{high}$ and $z_{low}$ represent the relative values of the open, high, low prices compared with the close price at current time step, respectively. $z_{close}$ and $z_{adj\_close}$ represent the relative values of the closing and adjusted closing prices compared with time step \(t-1\). $z_{dk}$ represents a long-term moving average of the adjusted close prices during the last \(k\) time steps compared to the current close price.  

\begin{table}[t]
\setlength\tabcolsep{4pt}
    \centering
    \begin{tabular}{|l|cccccc|}
    \hline
    \textbf{Dataset} & \textbf{Market} & \textbf{Freq} & \textbf{Stock} & \textbf{Days} & \textbf{From} & \textbf{To} \\ \hline
    ACL18 & US & 1d & 87 & 1258 & 12/09/01 & 17/09/01 \\
    SZ50 & China & 1h & 47 & 1036 & 16/06/01 & 20/09/01 \\
    \hline
    \end{tabular}
    \caption{Dataset statistics detailing market, data frequency, number of stocks, trading days and chronological period}
    \label{tab:dataset}
\end{table}

\begin{table}[t]
\setlength\tabcolsep{11pt}
    \newcommand{\tabincell}[2]{\begin{tabular}{@{}#1@{}}#2\end{tabular}}
    \centering
    \begin{tabular}{|l|l|}
    \hline
    \textbf{Features}  & \textbf{Calculation Formula}  \\ \hline
    $z_{open},z_{high},z_{low}$ & e.g., $ z_{open} = open_t/close_t - 1$ \\ 
    $z_{close}, z_{adj\_close}$ & e.g., $ z_{close} = close_t / close_{t-1} - 1$\\ \hline
    \tabincell{l}{$z_{d\_5}, z_{d\_10}, z_{d\_15}$ \\$z_{d\_20}, z_{d\_25}, z_{d\_30}$} & e.g., \(z_{d\_5} = \frac{{\textstyle \sum_{i=0}^{4}adj\_close_{t-i}/5}}{\textstyle adj\_close_{t}}-1\)  \\ \hline
    \end{tabular}
    \caption{Features to describe the stock markets with formulas}
    \label{tab:feature}
\end{table}


\subsection{Evaluation Metrics}
We evaluate AlphaMix on four different financial metrics including one profit criterion and three risk-adjusted profit criteria:
\begin{itemize}
    \item \textbf{Total Return (TR)} is the overall return rate of the whole trading period.
    It is defined as \( TR = \frac{n_{t}-n_{1}}{n_{1}} \), where \({n}_{t}\) is the final net value and \({n}_{1}\) is the initial net value.
    \item \textbf{Sharpe Ratio (SR)} considers the amount of extra return that a trader receives per unit of increase in risk. It is defined as: \( SR = {E}[\mathbf{r} ]/\sigma[\mathbf{r}] \), where \(\mathbf{r}\)  denotes the historical sequence of daily return rate.
    \item \textbf{Calmar Ratio (CR)} is defined as \(CR = \frac{E[\mathbf{r}]}{MDD}\). It is computed as the expected return divided by the Maximum Drawdown (MDD) of the whole trading period, where MDD measures the largest loss from any peak to show the worst case.
    \item \textbf{Sortino Ratio (SoR)} applies downside deviation (DD) as the risk measure. It is defined as: \(SoR = \frac{E[\mathbf{r}]}{DD}\), where DD is the variance of negative return. 
\end{itemize}

\subsection{Training Setup}
We perform all experiments on a Tesla V100 GPU. Grid search is applied to find the optimal hyperparameters based on validation total return for all methods. We explore the number of experts in range 2 to 8, coefficient \(\lambda\) from \(L_{individual}\) in list \([0.1, 0.5, 1, 5, 10]\), hidden states in range \([16, 32, 64]\). We use Adam as the optimizer with learning rate \(\alpha \in (1e^{-5}, 1e^{-3})\) and train AlphaMix for 10 epochs on all datasets. It takes 1.5 and 1.1 hours to train and test on ACL18 and SZ50 datasets, respectively. As for other baselines, we use the default settings in their public implementations\footnote{Qlib: https://github.com/microsoft/qlib}.

\subsection{Baseline}
We compare AlphaMix with 13 baseline models consisting of recurrent neural network (RNN), non-recurrent neural network (NRNN), boosting decision tree (BDT) and ensemble learning (ENS) methods:
\textit{Recurrent Neural Network (RNN)}
\begin{itemize}
    \item \textbf{LSTM-C \cite{nelson2017stock}} and \textbf{LSTM-R \cite{chen2015lstm}} use the vanilla LSTM network to predict stock movement and return rate, respectively. We apply a two-layer LSTM with hidden size 32. 
    \item \textbf{SFM \cite{zhang2017stock}} is a state frequency memory recurrent network, which decomposes prices into signals of different frequencies via Discrete Fourier Transform.
    \item \textbf{GRU-C \cite{shen2018deep}} and \textbf{GRU-R \cite{shen2018deep}} use a newer generation of recurrent networks with gated recurrent units for stock movement and return rate prediction, respectively. We apply a two-layer GRU with hidden size 64.
\end{itemize}
\textit{Non-Recurrent Neural Network (NRNN)}
\begin{itemize}
    \item \textbf{ALSTM \cite{qin2017dual}} adds an external attention layer into LSTM to attentively aggregate information from all hidden states in previous timestamps. 
    \item \textbf{MLP-C \cite{naeini2010stock}} and \textbf{MLP-R \cite{naeini2010stock}} apply multi-layer perceptron for stock movement and return prediction, respectively. We apply a three-layer MLP with hidden size 128.
    \item \textbf{GTrans \cite{ding2020hierarchical}} is a novel hierarchical Gaussian transformer-based model for stock prediction.
\end{itemize}
\textit{Boosting Decision Tree (BDT)}
\begin{itemize}
    \item \textbf{LightGBM \cite{ke2017lightgbm}} is an efficient implementation of gradient boosting decision tree with gradient-based one-side sampling and exclusive feature bundling.
    \item \textbf{CatBoost \cite{prokhorenkova2018catboost}} is a gradient boosting toolkit with ordered boosting and innovative categorical feature processing.
\end{itemize}
\textit{Ensemble Learning Methods (ENS)}
\begin{itemize}
    \item \textbf{EANN \cite{de2018designing}} is an ensemble model of deep neural networks with different architectures and training mechanisms.
    \item \textbf{DNNE \cite{yang2017stock}} is a bagging based ensemble model to efficiently train and combine multiple deep models.
\end{itemize}

\begin{table*}[!htb]

\centering
\resizebox{2.1\columnwidth}{!}{
\begin{tabular}{|c|c|c|cccc|cccc|}

\hline
\multicolumn{3}{ |c | }{} & \multicolumn{4}{ c | }{ACL18 (US)} & \multicolumn{4}{ c| }{SZ50 (China)}  
\\ 
\cline{1-11}
Type & Model & Task & TR(\%)$\uparrow$ & SR$\uparrow$ & CR$\uparrow$ & SoR$\uparrow$ & TR(\%)$\uparrow$ &  SR$\uparrow$ &  CR$\uparrow$ & SoR$\uparrow$  \\
\hline
&LSTM-C \cite{nelson2017stock} & Clf & 19.46 $\pm$ 4.29 & 1.48 $\pm$ 0.34 & 3.30 $\pm$ 1.14 & 2.29 $\pm$ 0.59 & 20.71 $\pm$ 7.75 & 1.01 $\pm$ 0.32 & 1.57 $\pm$ 0.52 & 1.28 $\pm$ 0.42 \\
&GRU-C \cite{shen2018deep} & Clf & 14.05 $\pm$ 7.85 & 1.06 $\pm$ 0.61 & 2.26 $\pm$ 1.91 & 1.51 $\pm$ 0.95 & 36.47 $\pm$ 27.97 & 1.25 $\pm$ 0.76 & 2.00 $\pm$ 1.35 & 1.74 $\pm$ 1.14 \\
RNN& SFM \cite{zhang2017stock} & Clf & 15.14 $\pm$ 2.99 & 1.25 $\pm$ 0.23 & 1.86 $\pm$ 0.47 & 1.93 $\pm$ 0.32 & 17.25 $\pm$ 2.34 & 0.79 $\pm$ 0.08 & 1.16 $\pm$ 0.14 & 1.04 $\pm$ 0.11 \\
&LSTM-R \cite{chen2015lstm} & Reg & 19.76 $\pm$ 5.06 & 1.53 $\pm$ 0.34 & 3.02 $\pm$ 1.34 & 2.15 $\pm$ 0.35 & 32.81 $\pm$ 21.98 & 1.20 $\pm$ 0.62 & 2.55 $\pm$ 1.76 & 1.74 $\pm$ 0.92 \\
&GRU-R \cite{shen2018deep} & Reg & 6.89 $\pm$ 7.65 & 0.55 $\pm$ 0.61 & 1.16 $\pm$ 1.26 & 0.79 $\pm$ 0.88 & \textcolor[RGB]{255,0,185}{$44.99^{\clubsuit} \pm$ 10.09} & 1.47 $\pm$ 0.24 & \textcolor[RGB]{255,0,185}{$3.18^{\clubsuit} \pm$ 0.52} & 2.29 $\pm$ 0.37 \\ \hline
&MLP-C \cite{naeini2010stock} & Clf & 22.05 $\pm$ 6.33 & 1.86 $\pm$ 0.57 & \textcolor[RGB]{255,0,185}{$3.69^{\clubsuit}$ $\pm$ 1.77} & 2.51 $\pm$ 0.87 & 43.48 $\pm$ 7.33 & 1.81 $\pm$ 0.14 & 2.87 $\pm$ 0.30 & 2.45 $\pm$ 0.23 \\
NRNN & GTrans \cite{ding2020hierarchical} & Clf & 16.45 $\pm$ 10.89 & 1.27 $\pm$ 0.82 & 3.07 $\pm$ 2.64 & 2.02 $\pm$ 1.37 & 16.79 $\pm$ 22.17 & 0.58 $\pm$ 0.77 & 0.85 $\pm$ 1.18 & 0.80 $\pm$ 1.07 \\
&MLP-R \cite{naeini2010stock} & Reg & 19.20 $\pm$ 5.44 & 1.52 $\pm$ 0.37 & 3.42 $\pm$ 1.86 & 2.14 $\pm$ 0.65 & 35.60 $\pm$ 17.98 & 1.30 $\pm$ 0.67 & $2.94^{\clubsuit} \pm$ 2.02 & 1.99 $\pm$ 1.05 \\
&ALSTM \cite{qin2017dual} & Reg & 15.13 $\pm$ 3.99 & 1.21 $\pm$ 0.32 & 2.63 $\pm$ 0.80 & 1.77 $\pm$ 0.49 & 32.30 $\pm$ 4.55 & 0.99 $\pm$ 0.13 & 1.70 $\pm$ 0.31 & 1.37 $\pm$ 0.26 \\
\hline

BDT & LGBM \cite{ke2017lightgbm} & Reg & 4.74 $\pm$ 3.11 & 0.36 $\pm$ 0.26 & 0.81 $\pm$ 0.32 & 0.56 $\pm$ 0.35 & 15.26 $\pm$ 3.11 & 0.66 $\pm$ 0.22 & 0.85 $\pm$ 0.23 & 0.85 $\pm$ 0.27 \\
&CatBoost \cite{prokhorenkova2018catboost} & Reg & 9.27 $\pm$ 3.27 & 0.70 $\pm$ 0.25 & 1.30 $\pm$ 0.55 & 1.00 $\pm$ 0.37 & 21.04 $\pm$ 1.81 & 1.06 $\pm$ 0.08 & 1.72 $\pm$ 0.23 & 1.42 $\pm$ 0.12 \\
\hline
ENS & EANN \cite{de2018designing} & Clf & \textcolor[RGB]{255,0,185}{23.32 $\pm$ 3.99} & \textcolor[RGB]{255,0,185}{1.97 $\pm$ 0.35} & 3.39 $\pm$ 0.91 & \textcolor[RGB]{255,0,185}{2.69 $\pm$ 0.60} & 44.30 $\pm$ 5.19& \textcolor[RGB]{255,0,185}{1.83 $\pm$ 0.14} & 2.89 $\pm$ 0.43 & \textcolor[RGB]{255,0,185}{2.51 $\pm$ 0.20}  \\
& DNNE \cite{yang2017stock} & Reg & 19.92 $\pm$ 5.17 & 1.60 $\pm$ 0.36 & $3.51^{\clubsuit}$ $\pm$ 1.73 & 2.26 $\pm$ 0.61 &40.74 $\pm$ 12.62&1.33 $\pm$ 0.34&2.88 $\pm$ 0.78&2.09 $\pm$ 0.54 \\
\hline
MoE &AlphaMix & Mtl & \textcolor[RGB]{129,0,121}{$28.15^{\clubsuit \heartsuit} \pm 3.31$} & \textcolor[RGB]{129,0,121}{\textbf{${2.53^{\clubsuit \heartsuit} \pm 0.33}$}} & \textcolor[RGB]{129,0,121}{\textbf{${3.78^{\clubsuit \heartsuit} \pm 0.45}$}} & \textcolor[RGB]{129,0,121}{\textbf{${3.66^{\clubsuit \heartsuit} \pm 0.64}$}} & \textcolor[RGB]{129,0,121}{\textbf{${57.60^{\clubsuit \heartsuit}\pm8.29}$}} & \textcolor[RGB]{129,0,121}{\textbf{${2.03^{\clubsuit \heartsuit}\pm0.25}$}} & \textcolor[RGB]{129,0,121}{\textbf{${3.88^{\clubsuit \heartsuit}\pm0.62}$}} & \textcolor[RGB]{129,0,121}{\textbf{${2.98^{\clubsuit \heartsuit}\pm0.30}$}}  \\ \hline
\multicolumn{3}{ |c | }{\% Improvement over SOTA} & \textcolor[RGB]{0,162,96}{$20.71\uparrow$} & \textcolor[RGB]{0,162,96}{$28.42\uparrow$} & \textcolor[RGB]{0,162,96}{$2.43\uparrow$} & \textcolor[RGB]{0,162,96}{$36.05\uparrow$} & \textcolor[RGB]{0,162,96}{$28.02\uparrow$} & \textcolor[RGB]{0,162,96}{$10.92\uparrow$} & \textcolor[RGB]{0,162,96}{$22.01\uparrow$} & \textcolor[RGB]{0,162,96}{$18.72\uparrow$} \\ \hline
\end{tabular}
}
\caption{Profitability comparison (mean and standard deviation of 10 individual runs) with 13 baselines including recurrent network (RNN), non-recurrent network (NRNN), boosting decision tree (BDT) and ensemble (ENS) methods with different task formulations (classification, regression or multi-task learning). Purple and pink show best \& second best results. $\heartsuit$ and $\clubsuit$ indicates improvement over SOTA baseline and EANN is statistically significant ($p<0.01$) under Wilcoxon's signed rank test.}
\label{fig:performance}
\end{table*}

\section{Results and Analysis}
\subsection{Profitability Comparison with Baselines}
We compare AlphaMix with 13 state-of-the-art baselines of various formulations in terms of four financial criteria in Table \ref{fig:performance}. We observe that AlphaMix consistently generates significantly (\(p<0.01\)) higher performance on all four criteria than all baselines across both US and China markets. In the US market, AlphaMix outperforms the second best by 20.71\%, 28.42\%, 2.43\%, 36.05\% in terms of TR, SR, CR, and SoR. As for the China market, AlphaMix outperforms the second best by 28.02\%, 10.92\%, 22.01\%, 18.72\% in terms of TR, SR, CR, and SoR. In general, ensemble learning methods perform better than single deep learning methods (RNN and NRNN), as they combine multiple submodels for more robust and profitable investment decisions. We postulate that both RNN and NRNN methods have the ability to capture the internal patterns of financial data since there is no obvious performance difference between RNN and NRNN methods. Boosting decision tree models tend to overfit to the noise in training data and generalize poorly in test data with the lowest profitability. In addition, we observe that EANN is the most powerful baseline with the second best performance on 5 out of 8 metrics. MLP-C achieves robust performance on both markets and GRU-R performs particularly well on the China market.

\begin{table*}[!htb]

\centering
\resizebox{2.1\columnwidth}{!}{
\begin{tabular}{|c|cccc|llll|llll|}

\hline
 & \multicolumn{4}{ c | }{} & \multicolumn{4}{ c | }{ACL18 (US)} & \multicolumn{4}{ c| }{SZ50 (China)}  
\\ 
\cline{1-13}
Models & Number of Experts & Multi-Task & Uncertainty & Router & TR(\%)$\uparrow$ & SR$\uparrow$ & CR$\uparrow$ & SoR$\uparrow$ & TR(\%)$\uparrow$ &  SR$\uparrow$ & CR$\uparrow$ & SoR$\uparrow$  \\ \hline
MLP-C & 1 &&& & 22.05 & 1.86 & 3.69 & 2.51  & 43.48  & 1.81  & 2.87  & 2.45  \\
EANN & 3 &&&& 23.32 & 1.97 & 3.39 & 2.69 & 44.30 & 1.83 & 2.89 & 2.51 \\\hline
& 1 & \(\surd\) & & & \textcolor[RGB]{218,56,51}{$22.73^*$}  & \textcolor[RGB]{218,56,51}{$1.86$}  & $4.17^{*\clubsuit}$  & \textcolor[RGB]{218,56,51}{$2.68^*$}  & $49.70^{*\clubsuit}$  & $1.89^{*\clubsuit}$  & $3.69^{*\clubsuit}$  & $2.89^{*\clubsuit}$  \\ 
 & 4 & & & & $25.58^{*\clubsuit}$ & $1.98^*$ & $4.55^{*\clubsuit}$ & $2.93^{*\clubsuit}$ & $45.58^*$ & \textcolor[RGB]{218,56,51}{1.69} & $2.95^{*}$ & $2.59^{*\clubsuit}$ \\ 
AlphaMix & 4 & \(\surd\) & & & $25.75^{*\clubsuit}$ & $2.08^{*\clubsuit}$ & $\textbf{4.79}^{*\clubsuit}$ & $3.11^{*\clubsuit}$ & $52.15^{*\clubsuit}$ & $1.86^*$ & $3.74^{*\clubsuit}$ & $2.96^{*\clubsuit}$ \\
 &4&\(\surd\)&\(\surd\)& & $26.71^{*\clubsuit}$ & $2.42^{*\clubsuit}$ & \textcolor[RGB]{218,56,51}{$3.29$} & $3.48^{*\clubsuit}$ & $57.05^{*\clubsuit}$ & $1.98^{*\clubsuit}$ & $3.52^{*\clubsuit}$ & $2.96^{*\clubsuit}$ \\
&4&\(\surd\)&\(\surd\)&\(\surd\)& $\textbf{28.15}^{*\clubsuit}$ & $\textbf{2.53}^{*\clubsuit}$ & $3.78^{*\clubsuit}$ & $\textbf{3.66}^{*\clubsuit}$ & $\textbf{57.60}^{*\clubsuit}$ & $\textbf{2.03}^{*\clubsuit}$ & $\textbf{3.88}^{*\clubsuit}$ & $\textbf{2.98}^{*\clubsuit}$ \\
\hline
\end{tabular}
}
\caption{Ablation studies over different AlphaMix's components (mean of 10 independent runs). \(\surd\) indicates adding the component to AlphaMix. Red indicates worse than either MLP-C or EANN. Bold shows the best results. $*$ and $\clubsuit$ indicate improvement over MLP-C and EANN, respectively, is statistically significant ($p<0.01$) under Wilcoxon's signed rank test.  }
\label{fig:ablation}
\end{table*}

\subsection{Model Component Ablation Study}
We conduct ablation studies on AlphaMix's investment profitability benefits from each of its components in Table \ref{fig:ablation}. First, we observe that the multi-task formulation can benefit profitability (22.05 to 22.73 and 43.48 to 49.70 on US and China markets respectively), which indicates jointly learning stock movement direction and return rate can lead to a more robust market representation. Next, we find that AlphaMix with 4 experts can outperform SOTA baseline in 7 out of 8 metrics, which validates that the mixture of experts is effective to achieve higher profitability. The performance consistently increases when the multi-task learning scheme is added. Furthermore, we observe that the uncertainty-aware loss significantly improves AlphaMix in terms of TR, SR, and SoR. As for CR, the performance slightly decreases due to the use of volatility as the risk measure and ignorance of the maximum drawdown. The future direction is to design new risk evaluation heuristics to improve CR. Finally, we add the routers to dynamically pick experts on an as-needed basis, which leads to the best performance on 7 out of 8 metrics across the two datasets. Comprehensive ablation studies demonstrate the effectiveness of all components in AlphaMix.



\subsection{AlphaMix as a Universal Framework}
\begin{figure}[!thb]
  \centering
    
     \subfloat{%
      \includegraphics[width=0.23\textwidth]{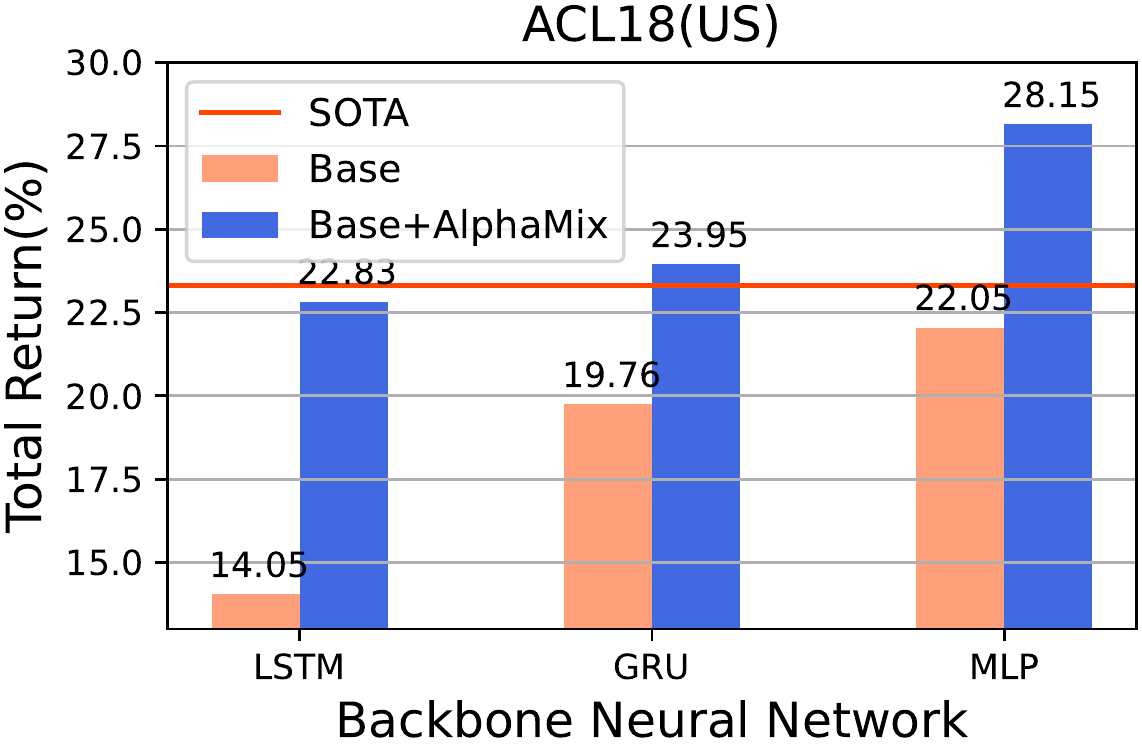}
     }
     \hfill
     \subfloat{%
      \includegraphics[width=0.23\textwidth]{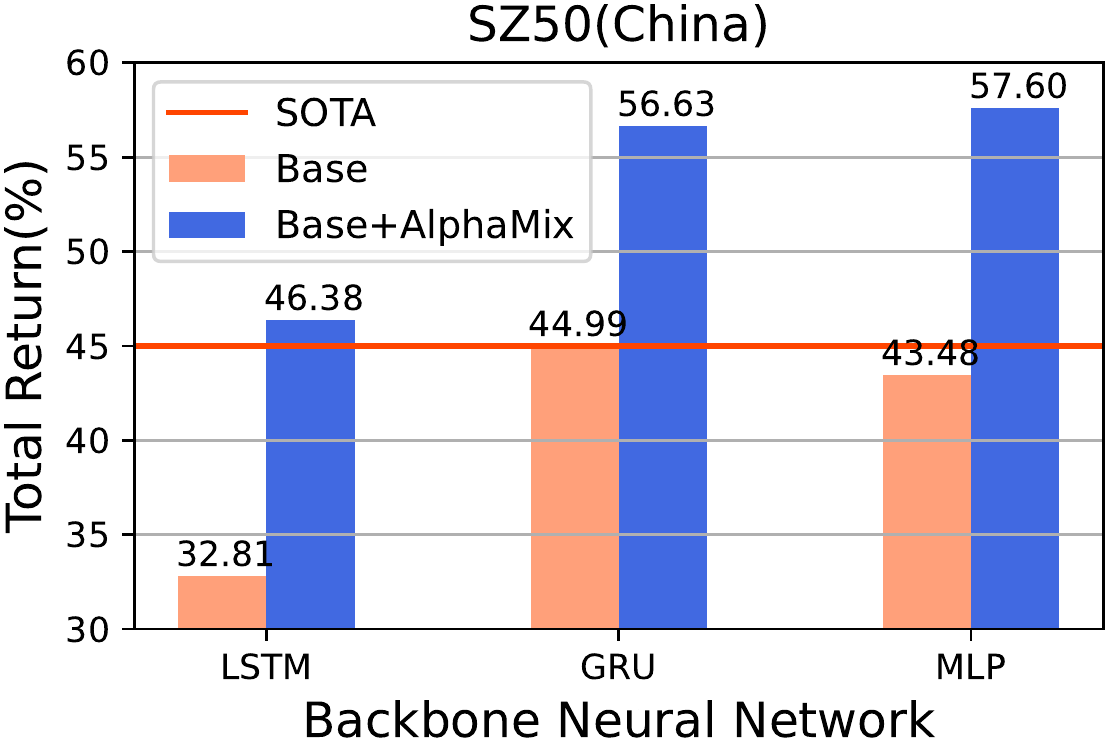}
     }

     \caption{Performance gains of AlphaMix as a universal framework with different backbone neural networks}
     \label{fig:universal}
\end{figure}

AlphaMix is a universal framework that can be applied to various backbone neural networks. We firstly train three basic deep learning models (LSTM, GRU, and MLP), then extend them with AlphaMix on both datasets, and report the performance gains in Figure \ref{fig:universal}. We observe that AlphaMix with different backbone networks can consistently outperform the previous SOTA (red line) in most cases (5 out of 6). Through combining different backbones with AlphaMix, there are at least 4.19\% and 11.64\% absolute total return improvements on US and China markets, respectively, which indicates the effectiveness of AlphaMix as a universal framework.  
\subsection{On the Effectiveness of Experts Router}

\begin{figure}[!thb]
  \centering
    
     \subfloat{%
      \includegraphics[width=0.23\textwidth]{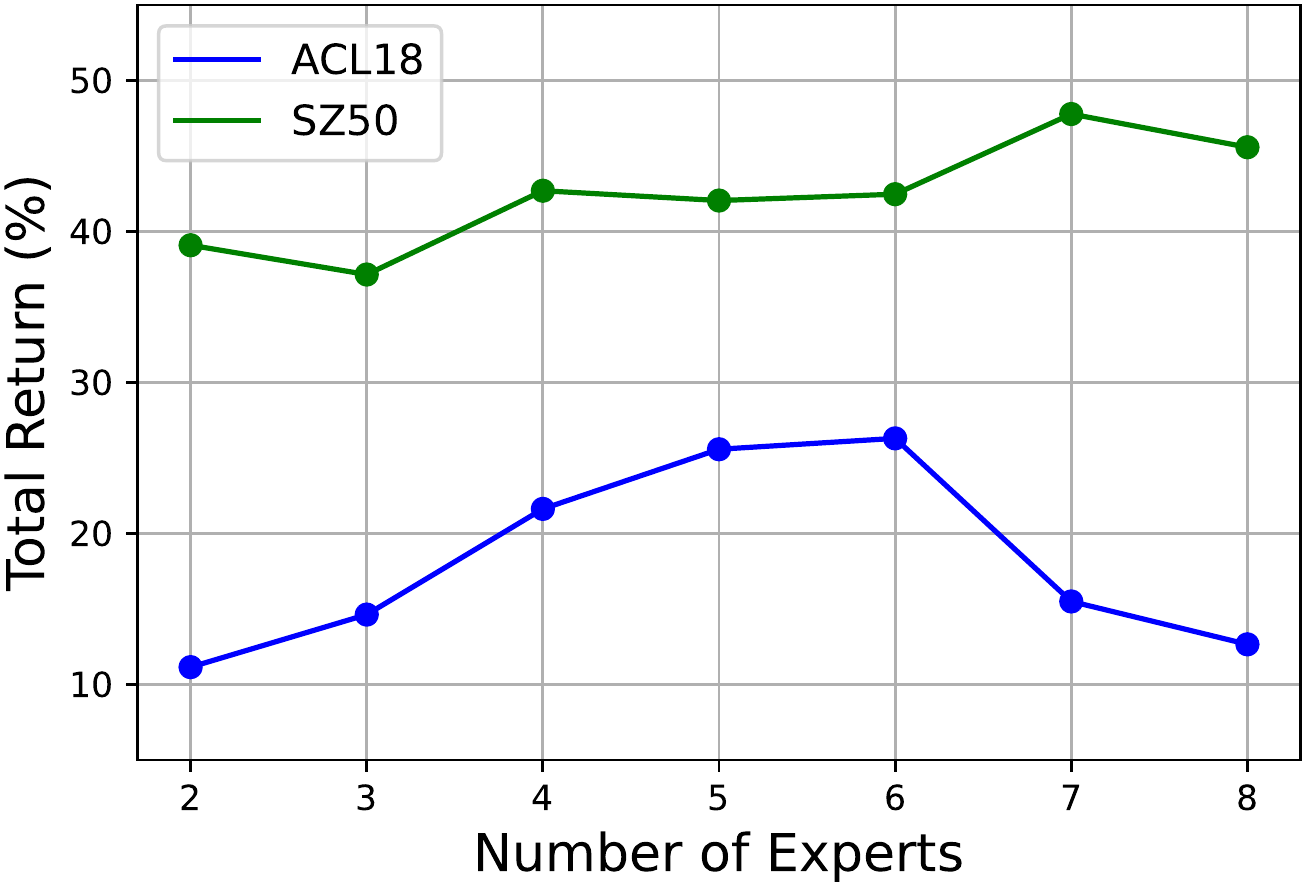}
     }
     \hfill
     \subfloat{%
      \includegraphics[width=0.23\textwidth]{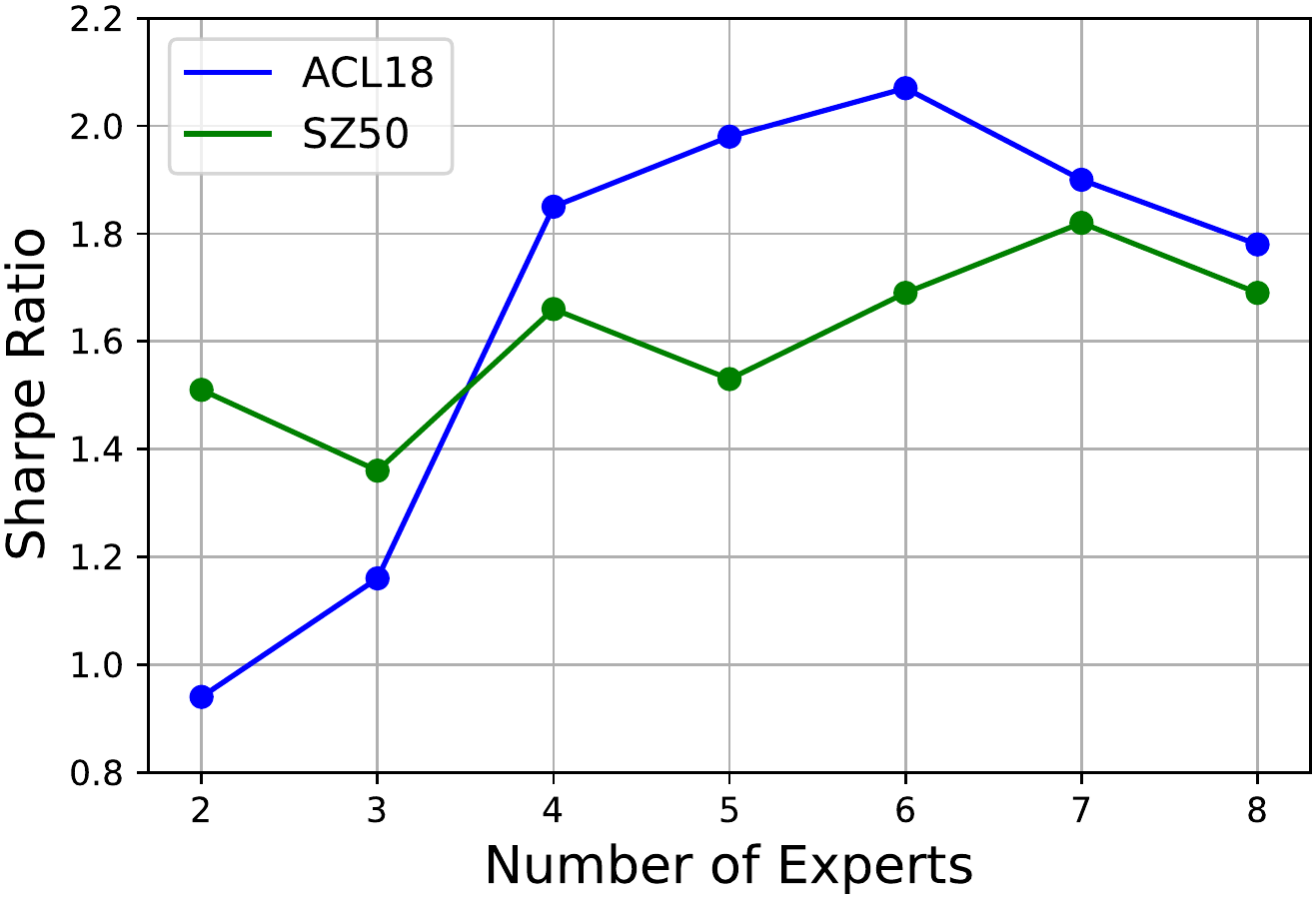}
     }
     \caption{The impact of the number of experts on AlphaMix in terms of TR and SR on ACL18 and SZ50}
     \label{fig:expert_num}
\end{figure}

\begin{figure}[!thb]
\begin{center}
\includegraphics[width=0.4\textwidth]{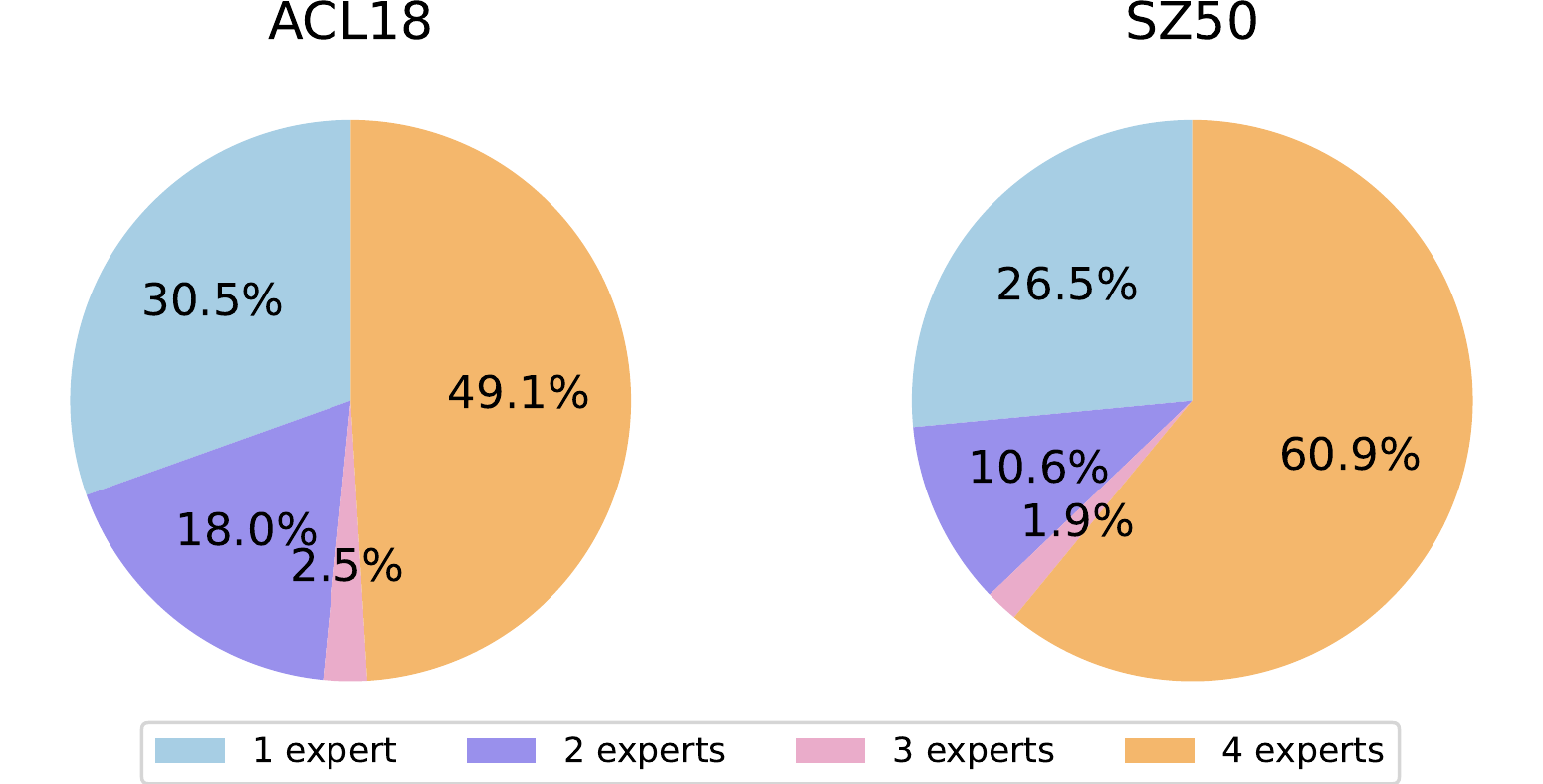}
\caption{The proportion of the number of experts allocated for AlphaMix with 4 experts on ACL18 and SZ50}
\label{fig:expert_pie}
\end{center}
\end{figure}

We analyze the effectiveness of the routers trained in Stage two from two perspectives. First, we explore the impact of the number of experts on AlphaMix's performance in Figure \ref{fig:expert_num}. With the increase of the number of experts, the performance gradually increase. AlphaMix achieves the best performance when \(k\) is 5 and 6 on US and China markets, respectively. When the number of experts is larger than 6, there is a smooth performance decay, which indicates that it is hard to make a consistent investment decision when there are too many opinions from different experts. The number of experts has a great impact on AlphaMix with a relative 135\% and 120\% improvement from the worst to the best in terms of TR and SR in the US market. As for the China market, it is less sensitive to the number of experts with a relative 22\% and 21\% improvement from the worst to the best in terms of TR and SR, respectively. 

Furthermore, we plot the number of experts allocated to each split of AlphaMix with 4 experts in Figure \ref{fig:expert_pie}. 30.5\% and 26.5\% of samples require one expert and 49.1\% and 60.9\% of samples require all four experts on ACL18 and SZ50, respectively. This shows that the developing market (e.g., China) is less efficient than the developed market (e.g., US) with more noise and usually needs more experts to make investment decisions. The percentage of samples where 2 and 3 experts are allocated is relatively small in both datasets. The possible reason is that the financial market is under stable or volatile conditions in most of time. There are less data samples belonging to the market conditions in between. In general, the routers are effective to control expert activation, which improve both the computational cost and performance of AlphaMix.

\begin{figure}[!thb]
  \centering
    
     \subfloat{%
      \includegraphics[width=0.23\textwidth]{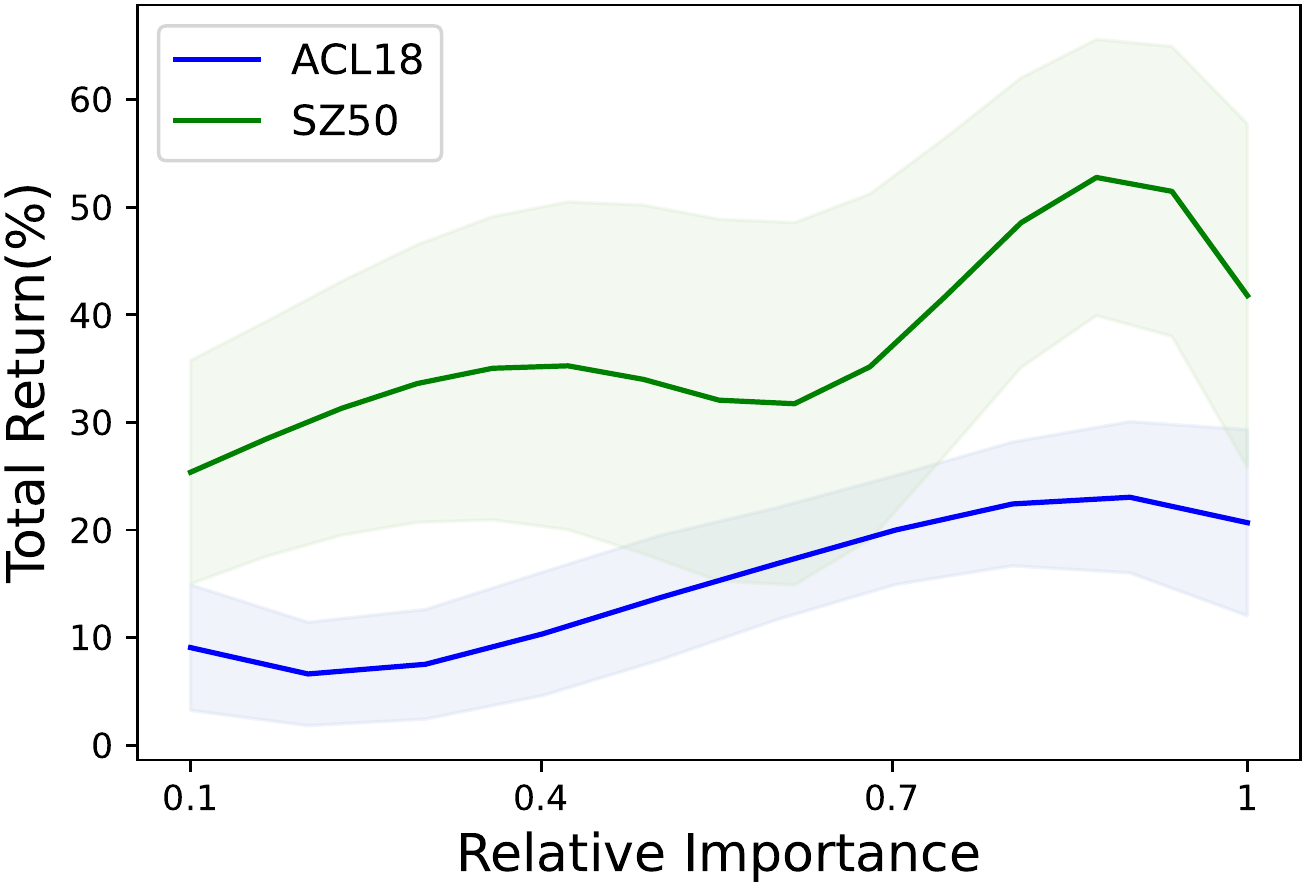}
     }
     \hfill
     \subfloat{%
      \includegraphics[width=0.23\textwidth]{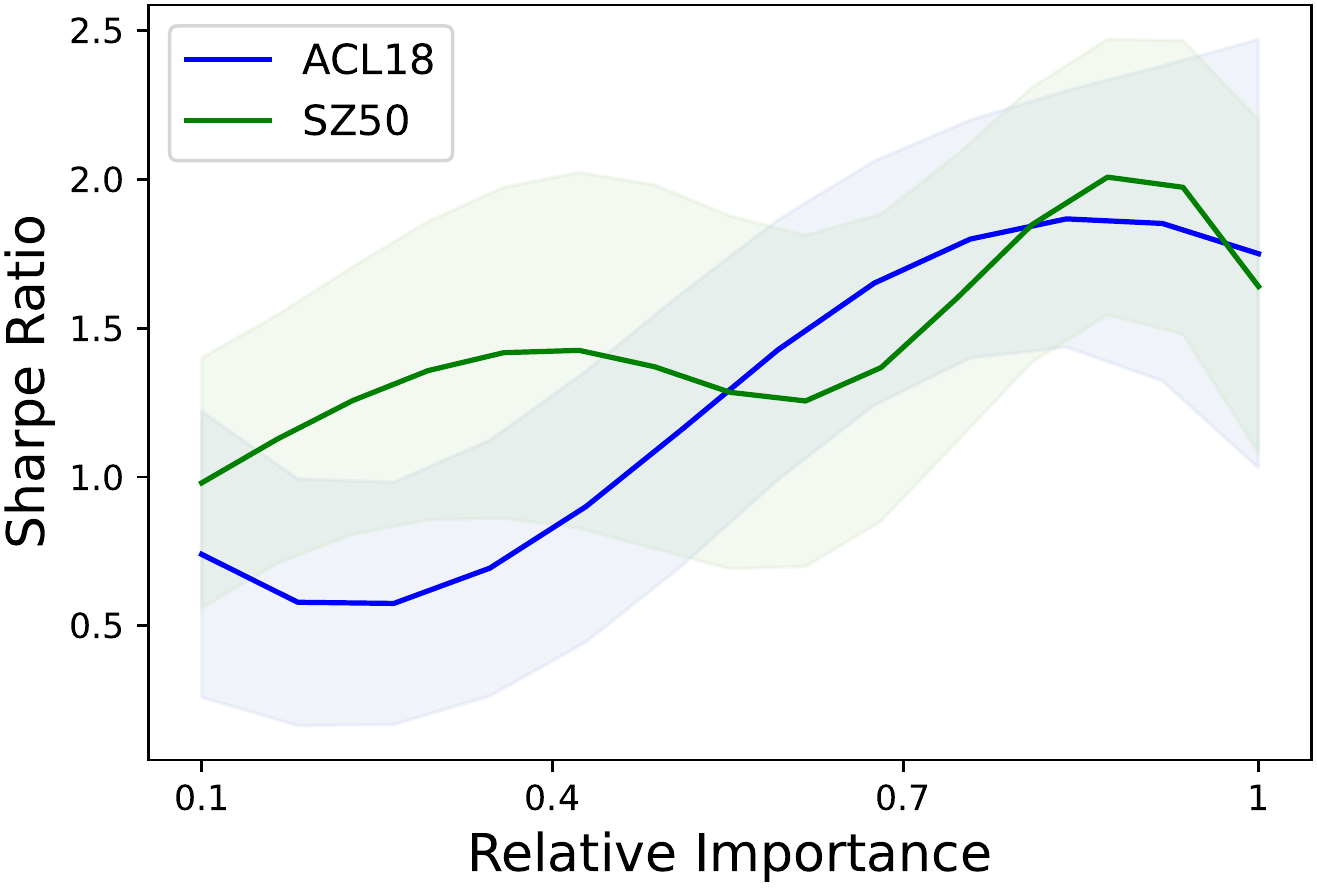}
     }

     \caption{Sensitivity to relative importance \(\lambda\) in terms of TR and SR on ACL18 and SZ50 }
     \label{fig:relative}
\end{figure}

\begin{figure}[!thb]
\begin{center}
\includegraphics[width=0.37\textwidth]{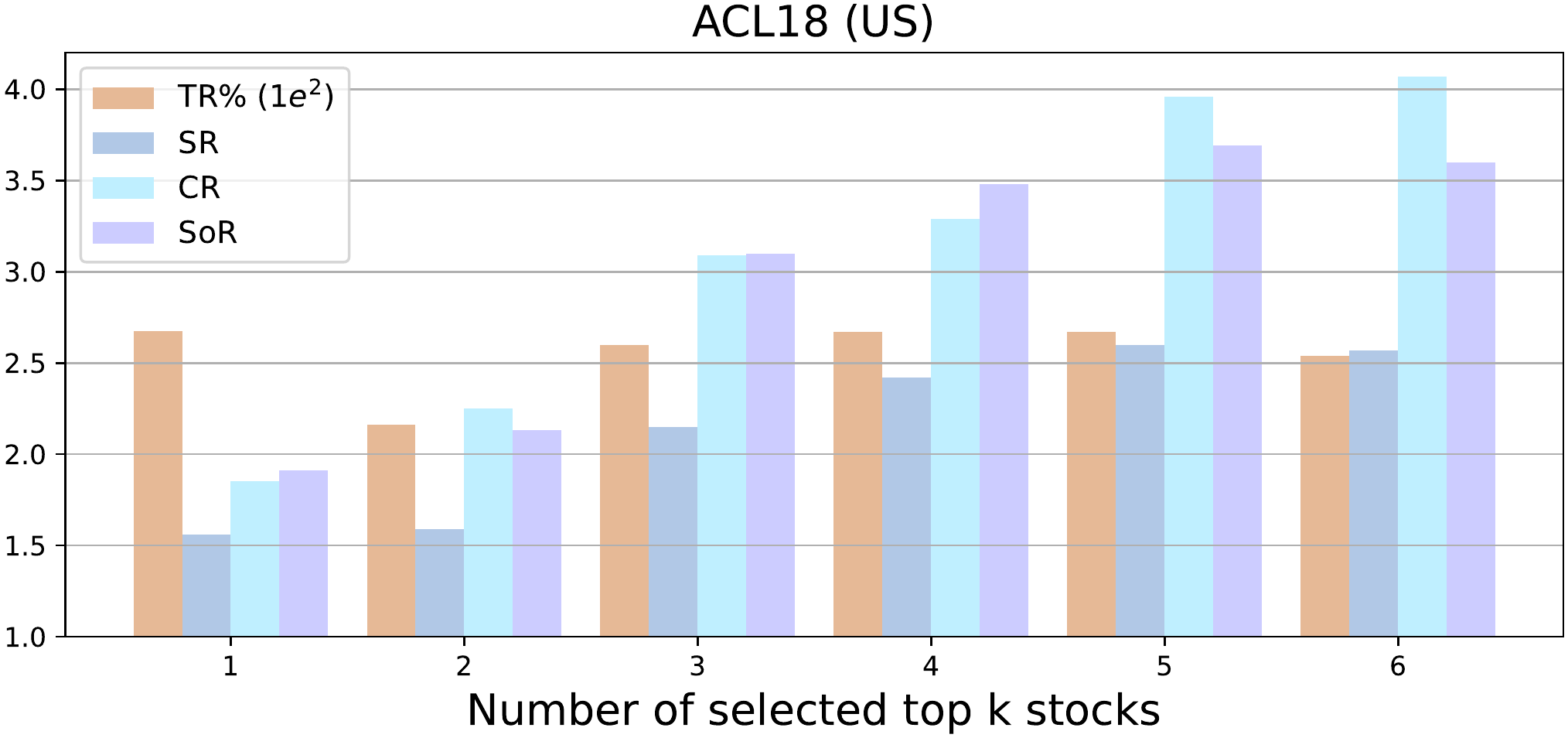}
\end{center}
\caption{Sensitivity to number of top stock selected \(k\)}
\label{fig:topk}
\end{figure}

\subsection{Parameter Analysis: Probing Sensitivity}
\quad \textit{Relative importance \(\lambda\) in the individual loss.} We study the performance change on AlphaMix with varying relative importance in Figure \ref{fig:relative}. We observe that both TR and SR increase smoothly and achieve the best performance when \(\lambda\) is around 0.85 on both datasets. The overall performance of AlphaMix is fairly robust over varying relative importance.

\textit{Number of selected top stocks k.} We analyze AlphaMix's profitability (TR, SR, SoR, CR) variation with the number of selected top k stocks on ACL18 in Figure \ref{fig:topk}. We find that AlphaMix performs consistently well in terms of TR. For the other three criteria, the performance gradually improves with the general best performance at \(k=5\), which shows AlphaMix's robustness and suitability to strategies with different risk taking appetites.

\section{Conclusion}
In this paper, we reformulate quantitative investment as a multi-task learning problem and propose a novel mixture-of-experts framework, AlphaMix, to develop profitable trading strategies. Specifically, we train multiple trading experts simultaneously on the individual uncertainty-aware loss. Later on, we train routers to dynamically allocate epxerts on an as-needed basis. Extensive experiments on both China and US stock markets demonstrate that AlphaMix significantly outperforms many strong baselines. Ablation studies show the effectiveness of the proposed components.

\bibliographystyle{ACM-Reference-Format}
\bibliography{reference}

\end{document}